\documentclass{aa}
\usepackage{pdflscape}
\usepackage{hyperref}
\usepackage{color}
\usepackage{graphicx}
\usepackage{times}
\usepackage{natbib}
\usepackage{amsmath,amssymb,amsfonts,txfonts}
\usepackage[colorinlistoftodos]{todonotes}

\definecolor{green}{rgb}{0,0.5,0}
\definecolor{grey}{rgb}{0.4,0.5,0.7}

\newcommand{\virgo}[1]{``#1''}
\newcommand{\der}{\mathrm{d}}

\newcommand{\rv}{r_{\Delta}}

\newcommand{\rtwo}{r_{200}}
\newcommand{\rfive}{r_{500}}

\newcommand{\gdm}{\gamma_{\rm{DM}}}
\newcommand{\rsdm}{r_{s}^{\rm{DM}}}
\newcommand{\rtwodm}{r_{200}^{\rm{DM}}}
\newcommand{\mtwodm}{M_{200}^{\rm{DM}}}

\newcommand{\rtwotot}{r_{200}^{\rm{tot}}}

\newcommand{\mtot}{M_{\rm{tot}}}
\newcommand{\mdm}{M_{\rm{DM}}}
\newcommand{\mbcg}{M_{\rm{BCG}}^*}
\newcommand{\mgas}{M_{\rm{gas}}}
\newcommand{\mgal}{M_{\rm{gal}}^*}

\newcommand{\br}{\beta(r)}
\newcommand{\nr}{\nu(r)}
\newcommand{\sigtr}{\sigma^2_r(r)}
\newcommand{\sr}{\sigma(r)}

\newcommand{\rmt}{r_{-2}}
\newcommand{\rs}{r_{s}}

\newcommand{\rj}{r_{\rm j}}

\newcommand{\rn}{r_{\nu}}

\newcommand{\ra}{r_{\beta}}
\newcommand{\binf}{\beta_\infty}

\newcommand{\likgal}{\mathcal{L}_{gal}}
\newcommand{\likbcg}{\mathcal{L}_{BCG}}

\newcommand{\cl}{AS1063}

\begin{document}


\institute{
INAF-Osservatorio Astronomico di Trieste,  via G. B. Tiepolo 11, I-34131, 
Trieste, Italy (e-mail: barbara.sartoris@inaf.it)
\label{oats}  \and
IFPU - Institute for Fundamental Physics of the Universe,  via Beirut 2, I-34014, 
Trieste, Italy 
\label{ifpu}  \and
Dipartimento di Fisica e Scienze della Terra, Universit\`a di Ferrara, Via Saragat 1, 44122 Ferrara, Italy
\label{unife}  \and
INAF - OAS, Osservatorio di Astrofisica e Scienza dello Spazio di Bologna, via Gobetti 93/3, I-40129 Bologna, Italy
\label{inafbo} \and
INFN, Sezione di Bologna, viale Berti Pichat 6/2, I-40127 Bologna, Italy
\label{infnbo} \and
INAF-Osservatorio Astronomico di Capodimonte, Via Moiariello 16, 80131 Napoli, Italy
\label{oac}  \and
Dipartimento di Fisica, Universit\`a  degli Studi di Milano, via Celoria 16, I-20133 Milano, Italy\label{unimilano} \and
DARK, Niels-Bohr Institute, Lyngbyvej 2, 2100 Copenhagen, Denmark
\label{dark}  \and
Academia Sinica Institute of Astronomy and Astrophysics (ASIAA), No. 1, Section 4, Roosevelt Road, Taipei 10617, Taiwan
\label{asiaa} \and
Kapteyn Astronomical Institute, University of Groningen, Postbus 800, 9700 AV Groningen, The Netherlands 
\label{Kapteyn}  \and
Dipartimento di Fisica, Univ. degli Studi di Trieste, via Tiepolo 11, I-34143 
Trieste, Italy
\label{units}  
}

\title{CLASH-VLT: a full dynamical reconstruction of the mass profile of Abell S1063 from 1 kpc out to the virial radius}
\titlerunning{\cl\ mass profile}

\author{
B. Sartoris\inst{\ref{oats},\ref{ifpu}}\and
A. Biviano\inst{\ref{oats},\ref{ifpu}}\and
P. Rosati\inst{\ref{unife},\ref{inafbo}}\and 
A. Mercurio\inst{\ref{oac}}\and 
C.~Grillo  \inst{\ref{unimilano},\,\ref{dark}}\and
S. Ettori \inst{\ref{inafbo},\ref{infnbo}}\and
M. Nonino \inst{\ref{oats}}\and
K. Umetsu \inst{\ref{asiaa}}\and
P. Bergamini \inst{\ref{inafbo}}\and
G.~B.~Caminha \inst{\ref{Kapteyn}}\and
M. Girardi \inst{\ref{units}} 
}
\authorrunning{Sartoris et al.}

\abstract
{The shape of the mass density profiles of cosmological halos informs us of the nature of dark matter (DM) and DM-baryons interactions. Previous estimates of the inner slope of the mass density profiles of clusters of galaxies are in opposition to predictions derived from numerical simulations of cold dark matter (CDM).}
{We determine the inner slope of the DM density profile of a massive cluster of galaxies, Abell S1063 (RXC J2248.7$-$4431) at $z=0.35$, with a dynamical analysis based on an extensive spectroscopic  campaign carried out with the VIMOS and MUSE spectrographs at the ESO VLT. This new data set provides an unprecedented sample of 1234 spectroscopic members, 104 of which are located in the cluster core ($R\lesssim 200$ kpc), extracted from the MUSE integral field spectroscopy. The latter also allows the  stellar velocity dispersion profile of the brightest cluster galaxy (BCG) to be measured out to 40 kpc.}
{We used an upgraded version of the MAMPOSSt technique to perform a joint maximum likelihood fit to the velocity dispersion profile of the BCG and to the velocity distribution of cluster member galaxies over a radial range from 1 kpc to the virial radius ($r_{200}\approx 2.7$ Mpc).}
{We find a value of $\gdm =0.99 \pm 0.04$ for the inner logarithmic slope of the DM density profile after marginalizing over all the other parameters of the mass and velocity anisotropy models. Moreover, the newly determined dynamical mass profile is found to be in excellent agreement with the mass density profiles obtained from the independent X-ray hydrostatic analysis based on deep Chandra data, as well as the strong and weak lensing analyses.}
{Our value of the inner logarithmic slope of the DM density profile $\gdm$ is in very good agreement with predictions from cosmological CDM simulations. 
We will extend our analysis to more clusters in future works. If confirmed on a larger cluster sample, our result makes this DM model more appealing than alternative models. } 
 
\keywords{Galaxies: clusters: general / Galaxies: kinematics and dynamics / dark matter / galaxies: clusters: individual: Abell S1063 / galaxies: clusters: individual: RXC J2248.7$-$4431}

\maketitle jikhiugil  fbmfgkljvemrfgvsrgmv

\section{Introduction} \label{s:intro}

A strong prediction of the cold dark matter (CDM) cosmological simulations is the existence of a universal shape for the dark matter (DM) mass density profile $\rho(r)$ of cosmological halos. In particular, in their seminal papers, \citet{NFW96,NFW97} suggested that the following Navarro–Frenk–White function (hereafter NFW),
\begin{equation}
  \rho(r) \propto (r/\rmt)^{-1} \times (1+r/\rmt)^{-2},
\end{equation}
adequately characterizes halos extracted from DM-only cosmological simulations over a wide range of masses, from the center, $r=0$, to the virial radius. The NFW profile is characterized by a logarithmic slope $\der \log \rho / \der \log r = -1$ for $r \rightarrow 0$, and $-3$ at large radii, and by a characteristic radius, $\rmt$, where $\der \log \rho / \der \log r=-2$.

With the development of more accurate simulations, the universal NFW shape has been questioned \citep[e.g.,][]{RPV07,DelPopolo10}. In particular, its inner slope might not be fixed, but continuously changing with $r$, becoming shallower towards the center \citep[e.g.,][]{Navarro+04,Stadel+09}. Moreover, the inner structure of halos can be influenced by collisional processes that affect the baryonic components of these halos. Central condensation of cooled gas \citep[e.g.,][]{Blumenthal+86,Peirani+17} 
would steepen the central $\rho(r)$ slope, while dynamical friction could have the opposite effect, transferring energy from infalling satellite galaxies to the central DM cusp \citep[e.g.,][]{EZSH01,DelPopolo09}. Other baryonic effects capable of reducing the central mass concentration are supernovae feedback in low-mass galaxies \citep[e.g.,][]{Governato+10} and  active galactic nuclei (AGN) feedback in clusters of galaxies \citep[e.g.,][]{MTMW12}. The combination and the net effect of these baryonic processes on the DM density profile is still under debate \citep[e.g.,][]{Schaller+15,LW15,Peirani+17,He19}.

The properties of DM itself can affect the inner slope of a halo $\rho(r)$. Warm, fuzzy, decaying, and self-interacting DM have been shown in numerical simulations to produce flat inner $\rho(r)$ slopes \citep{HBG00,BOT01,PMK10}. For example, the self-interacting DM models predict an inner halo density profile shallower than the NFW  \citep[e.g.,][]{SS00,Rocha13}.

Constraining the shape of $\rho(r)$ of cosmological halos can therefore provide valuable constraints on the properties of DM, the interplay between the DM and baryonic contents, and on the halos assembly process. Among cosmological halos, clusters of galaxies are the only ones for which $\rho(r)$ can be determined over a wide range of scales, using several observational tools, sometimes in combination. In particular, X-ray observations of the hot intra-cluster medium (ICM hereafter) can be used to probe the cluster mass distribution under the assumption of hydrostatic equilibrium, but typically only out to $\rfive$ \citep[e.g.,][]{ettori13xmass,giles17}, and with some systematic uncertainties due to the deviation from hydrostatic equilibrium in the outer regions \citep{Rasia+06} and due to the presence of cool components and temperature fluctuations near the cluster center \citep{ABA04}.  The Sunyaev-Zeldovich effect \citep[][for a recent review]{SZ70b,mroczkowski19}, due to the inverse Compton scattering of the cosmic microwave background (CMB) photons off the electrons of the ICM, can be used in combination with X-ray observations to probe $\rho(r)$ beyond $\rfive$, and up to $\rtwo$, as originally suggested by \cite{ameglio09}, and recently probed by \cite{shitanishi18} on a sample of objects observed with {\it Bolocam} and {\it Chandra} and by \cite{ettori19} on the X-COP sample of massive nearby galaxy clusters resolved with {\it Planck} and {\it XMM-Newton} exposures. Gravitational lensing is the most direct way of determining the (projected) total mass distribution in clusters within the strong regime within the central $\sim 300$ kpc \citep[e.g.][]{mellier93,zitrin12b,caminha16}, and within the weak regime outside the central region \citep[e.g.][]{Squires+96,umetsu14}.

The distribution of cluster members in projected phase-space can be used to trace the gravitational potential from the central $\sim50$ kpc to very large radii \citep[e.g.][]{RD06,biviano13}, using methods based on the Jeans equation for gravitational equilibrium \citep{binney87,Wojtak+09,MBB13}, or the so-called \virgo{caustic} method \citep{DG97,diaferio99}.  Finally, the stellar kinematics of the brightest cluster galaxy (BCG) provides an additional measure of the total cluster mass at very small radii, out to $\sim 100$ kpc \citep[e.g][]{dressler79,kelson02}. Whatever the method used to constrain $\rho(r)$ in clusters, it is important to distinguish the DM contribution from that due to baryons. For example, ignoring the baryonic contribution of the BCG could lead to over-estimate the inner slope of the DM $\rho(r)$. The baryonic contribution of the other galaxies is instead quite negligible, as most of the baryons outside the very cluster center are contributed by the ICM \citep[e.g.,][]{bivsal06,eckert19}.
 
 Previous observational determinations of the central slope of the DM $\rho(r)$ were based on a combination of strong lensing and the BCG kinematics, and found $\gdm<1$ with a large statistical significance, that is a profile flatter than NFW  \citep{STE02,sand04,sand08,newman09,newman11,newman13a,newman13b}. This finding stimulated discussion on the physical reason for this difference, be it the nature of DM or the interplay between baryons and DM \citep[see,   e.g.,][for a detailed discussion on this topic]{newman13b,He19}.
 
In this paper, we determine the mass profile of the cluster Abell S1063 (hereafter \cl) from $\simeq 1$ kpc to the cluster virial radius. \cl\ is a rich galaxy cluster at $z=0.3458$, also known as RXC J2248.7$-$4431 and it was first identified in \citet{ACO89}. \cl\  was observed with HST as part of the \textit{Cluster Lensing And Supernova  survey with Hubble} \citep[CLASH,][]{postman12}, and the  \textit{Frontier Fields} program \citep{lotz17}. An extensive spectroscopic campaign was conducted with VIMOS at the VLT as part of the CLASH-VLT ESO Large Programme \citep[ID 186.A-0798, P.I. P.Rosati,][Rosati et al. in prep]{rosati14}.  This data set has been complemented with observations obtained with the integral field spectrograph MUSE at the VLT to better probe the cluster core and BCG kinematics.

With respect to previous works \citep[e.g.][]{sand04,newman13a}, we performed a full dynamical analysis  of the potential well of the cluster using a combination off two different independent tracers: the kinematics of the cluster galaxy members and the stellar  velocity dispersion profile of the BCG. We deconvolved the different contributions to the cluster total mass profile, coming from the stellar mass of member galaxies, the hot gas component, the BCG stellar mass, and the DM.  Finally we compared the total mass profile obtained from the dynamical analysis with that obtained by using the hydrostatic mass from X-ray data, and a combination of strong and weak lensing mass measurements. 

The paper is structured as follows: in Section \ref{s:data}, we present our VIMOS and MUSE data sets and in Section \ref{s:mod} we describe our methodology. In Section \ref{s:res}, we provide the results on the mass profile of \cl; specifically, we derive the best-fit parameters of the DM mass profile, we show the contribution of the different matter components to the total cluster mass profile, and we compare the dynamical total mass profiles with those derived from other independent analyses. Finally, in Section \ref{s:disc}, we discuss our results and in Section \ref{s:conc}, we draw our conclusions.  

Throughout this paper, we adopt  $H_0=70$ km~s$^{-1}$~Mpc$^{-1}$, $\Omega_m=0.3$, $\Omega_\Lambda=0.7$. At the cluster redshift, 1 arcmin corresponds to 296 kpc.  Magnitudes are in the AB system. We call $\rv$ the radius that encloses an average density $\Delta$ times the critical density at the halo redshift.

\section{Data set}\label{s:data}

This work is based on the spectroscopic observations carried out with the VIMOS and the MUSE spectrographs at the VLT. The VIMOS CLASH-VLT spectroscopic campaign for \cl\ is presented in Mercurio et al. 2020 (in prep., hereafter M20). This yielded 3607 reliable redshifts measured over a field of $25\times 25$ arcmin$^2$, 1109 of which are classified as cluster members based on estimators which use the projected phase-space distribution of galaxies around the median redshift of the cluster $z=0.3458$ (see M20). 

The CLASH-VLT campaign was significantly enhanced and complemented in the cluster core using observations with the integral field spectrograph MUSE \citep{bacon10}. The MUSE data set consists of two pointings covering the NE and SW sides of the cluster core. Data reduction procedures and redshift measurements are fully described in \citet{karman15}, which presents the SW pointing obtained during MUSE science verification for a total exposure of 3.1 h (ID 60.A-9345, P.I. Caputi\&Grillo). The extension to the NE pointing (095.A-0653, P.I. Caputi), with a co-added exposure of 4.8 h, is described in \citet{karman17}. The seeing was measured to be $\sim 1.1$ and $0.9$ arcsec in the SW and NE pointings, respectively. The resulting catalogue of cluster members and multiply lensed images in the $\sim 2\times 2$ arcmin$^2$ central region was presented in  \citet{caminha16}, along with the first strong lensing model. The MUSE data provided 175 secure redshifts in addition to the 3607 aforementioned VIMOS campaign, 104 of which are classified as cluster galaxies.

\begin{figure*}
\hbox{\includegraphics[width=0.70\textwidth]{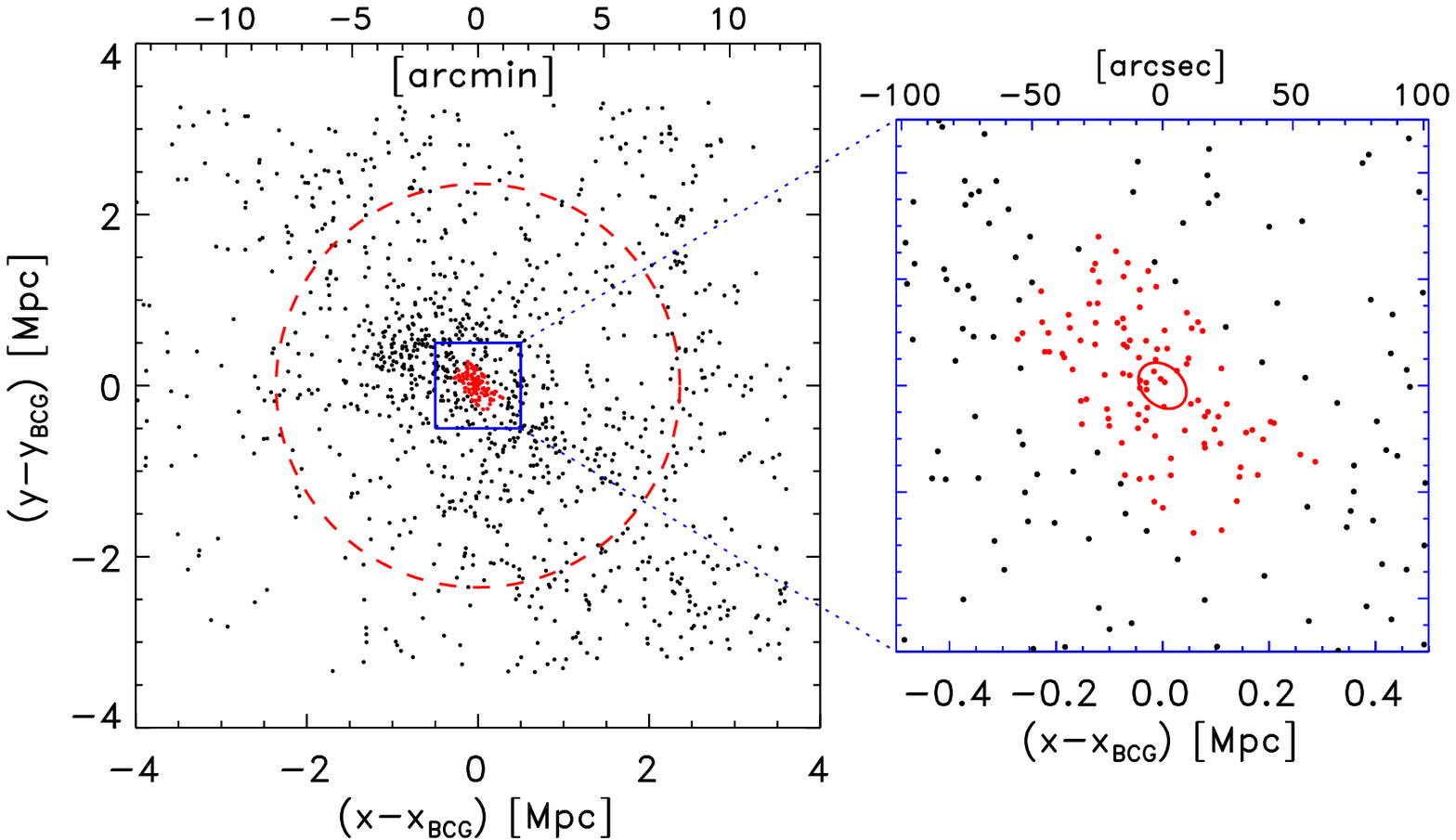}
\raisebox{10mm}{\includegraphics[width=0.30\textwidth]{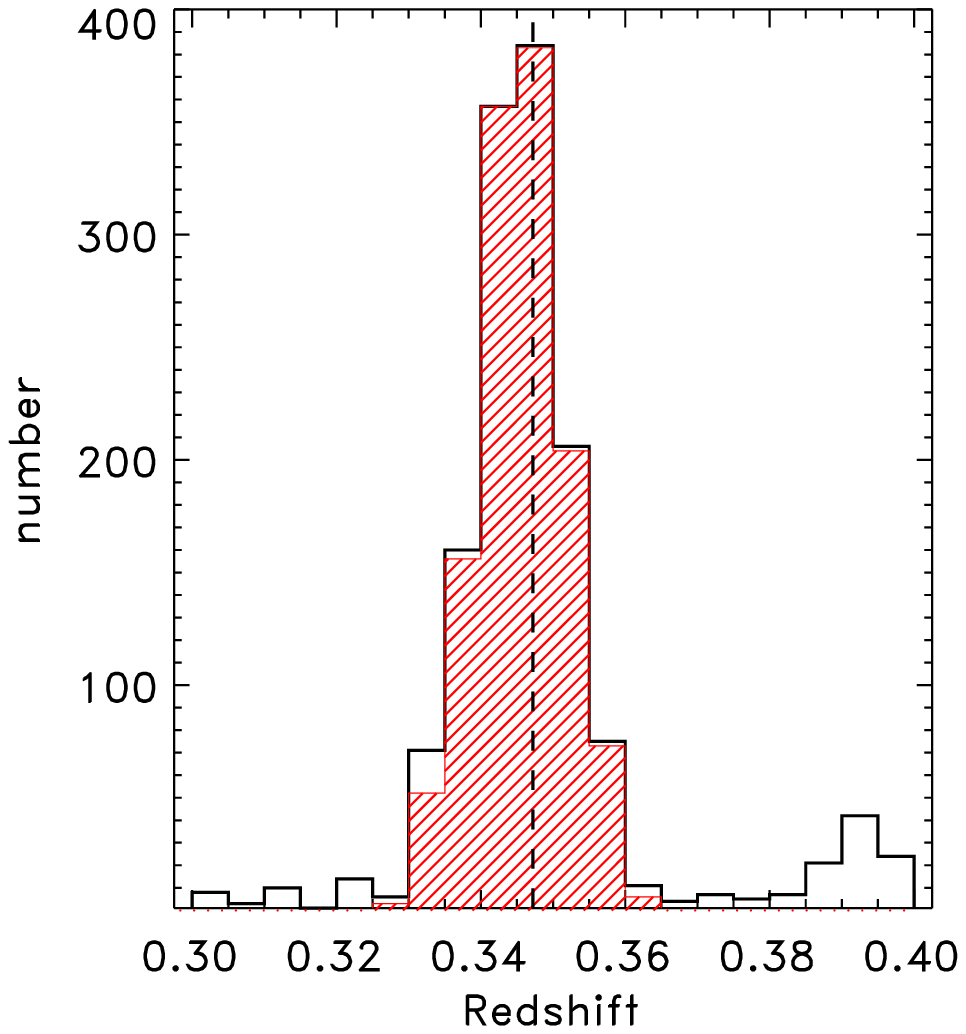}}
}\hspace*{-2cm}
\caption{Spatial distribution of the 1234 spectroscopic cluster members over the entire surveyed field (left). Galaxies with MUSE redshifts are indicated in red.  The red dashed circle represents $r_{200}=2.36$ Mpc from the lensing analysis of \citet{umetsu16}. The central Mpc (blue box) is enlarged in the central panel, where the ellipse indicates the BCG region within which its internal kinematics can be measured with MUSE (semi-major axis $= 50$ kpc, see Fig.\ref{fig:sigmaprof}). Right panel: redshift distribution of galaxies selected as cluster members (red histogram); the vertical dashed line indicates the BCG redshift ($z=0.3472$).}
\label{fig:xy}
\end{figure*}

By considering 21 additional redshifts from \citep{gomez12}, 30 from the Grism Lens-Amplified Survey from Space \citep{treu15}), and 17 unpublished redshifts from Magellan observations (D. Kelson, private communication), a total of 3850 spectroscopic redshift were analyzed and  1234 cluster members were selected for the dynamical analysis presented in this work (see M20). The spatial distribution of cluster galaxies and their redshift distribution are shown in Figure ~\ref{fig:xy}.

The radially dependent completeness of our spectroscopic sample is discussed in M20. It is essentially 100\% in the MUSE region, that is, $ 0< R\, {\rm (Mpc)} < 0.25$ Mpc, $\sim 80$\% in the range $0.25 < R\,{\rm (Mpc)}<1 $, and drops to $\sim 75 \%$ at $1< R\,{\rm (Mpc)} \le 2.75$ up to R=23.0 mag, with typical errors of $\sim 5$\%.

Repeated measurements of the same spectra were used to  estimate the average error on the radial velocities: 75 (153) km s$^{-1}$ for the spectra observed with the MR (LR, respectively) grism. As for the MUSE velocities, the average estimated error is 15 km s$^{-1}$. As for the velocities taken from the literature, we use the errors quoted in the reference papers.
\newline

\begin{figure*}
\begin{center}
 \includegraphics[width=0.80\textwidth]{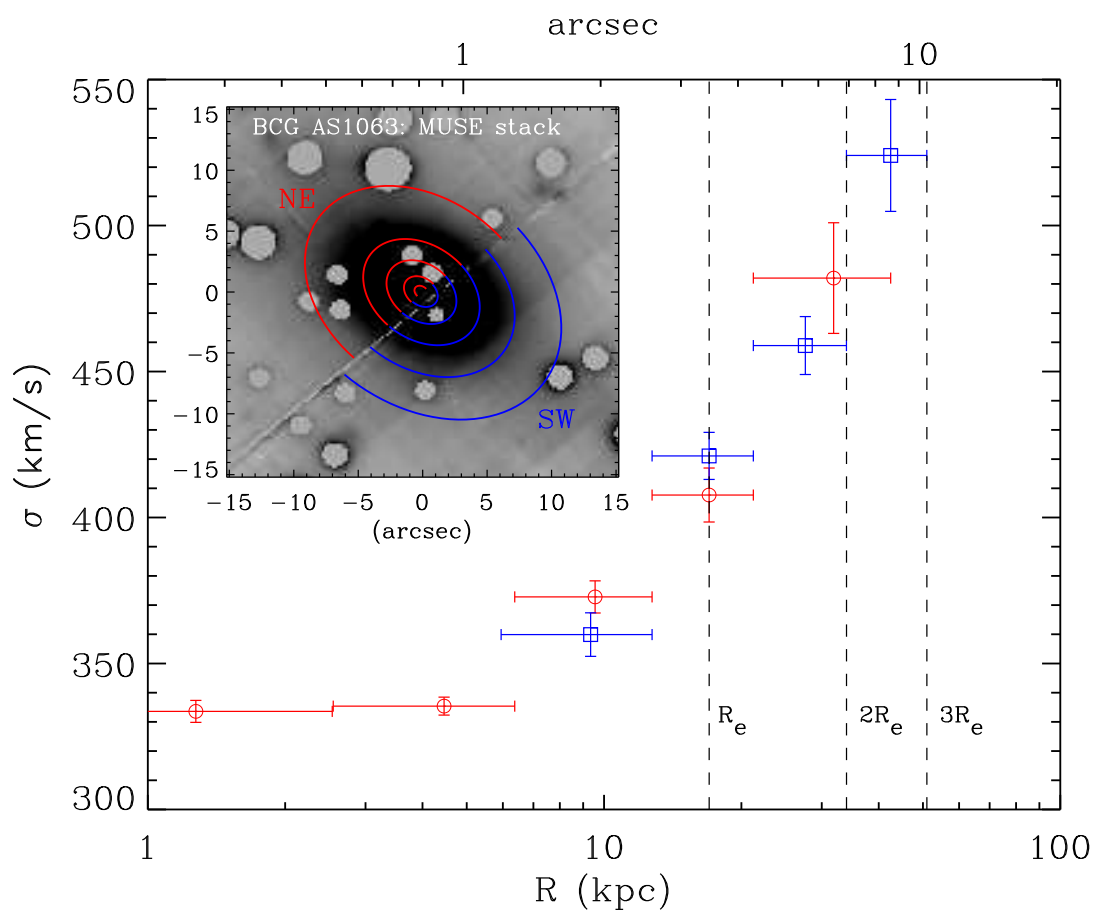}
\end{center}

\caption{LOS velocity dispersion profiles from the stellar kinematics of the BCG, obtained by extracting spectra in elliptical radial regions from the MUSE data cubes, independently in the NE (red circles) and SW (blue squares) portion of the BCG, as shown in the inset. The latter is a median stack of the MUSE data in the inner 30 arcsec around the BCG, with projected faint cluster members masked out in the spectral extraction. We note that the innermost radial bin (blue) in the SW side of the BCG is excluded since it contains the gap between the two MUSE pointings. Vertical dashed lines indicate 1,2,3 effective radii ($R_e=17$ kpc).}
\label{fig:sigmaprof}
\end{figure*}

The MUSE integral field data are also ideal to measure the spatially dependent internal kinematics of the BCG, with a spatial resolution limited by the seeing (sampled with pixels of 0.2 arcsec) and a velocity resolution of $\sim 50$ km s$^{-1}$. The stellar line-of-sight velocity dispersion (LOSVD) of the BCG in different radial bins is measured using the pPXF public software described in \citet{cappellari04}, recently upgraded and improved \citep{cappellari17}. The second moment of the line absorption profiles, that is, the width $\sigma$ of a Gaussian shape, is measured by convolving a set of stellar spectral templates to fit the MUSE spectra, taking into account the spectrograph resolution (2.6\AA). When preparing input spectra for pPXF, only the spectral region [4860--7160]\AA\ is considered, where the MUSE sensitivity is higher,
while narrow spectral regions with high variance, primarily in the
vicinity of sky and telluric lines are masked out.  The stellar
template library includes 105 spectra extracted from the NOAO
high-resolution spectral library \citep{gunn83,jacoby84}, with an intrinsic resolution
of 1.35\AA\ and a dispersion of 0.4\AA/pixel. Most of the stars have
the spectral type GKM, including also 10\% of A-stars, to broadly match
the underlying stellar populations of an ETG. The robustness of the
LOSVD measurements has been tested with extensive simulations
\citep{bergamini19} at varying signal-to-noise and
LOSVD, thus proving a realistic model for statistical and systematic
errors. The MUSE data on the BCG are divided into two halves in the two
pointings, approximately along the major axis of the BCG (see the inset in
Figure ~\ref{fig:sigmaprof}). While this poses some challenges which require masking
high-variance pixel regions near the edge of the MUSE fields, it does
offer an independent measurement of the $\sigma$-profile in elliptical
regions on each side of the BCG, where relatively faint cluster
galaxies projected in the inner core are masked out. In Figure~\ref{fig:sigmaprof}, we show
the $\sigma$-measurements in five radial elliptical bins ($b/a=0.75$, PA$=319^\circ $) in the NE
portion and 4 bins in the SW portion of the BCG, as a function of the
circularized projected radius.  
We note that the BCG light profile peaks in the NE side, whereas the central bin in the SW side is affected by the pointing edge. Therefore, the innermost $\sigma$ measurements within 10 kpc are extracted from the NE side. Signal-to-noise ratios (S/N) of the spectra (the average value over the aforementioned wavelength region) range from
$\sim 80$ in the inner radial bins to $\sim 24$ in the outer bins, which extend to
$\sim 3$ times the value of the effective radius in the SW side. 
These two independent
measurements of the $\sigma$-profile are in excellent agreement and
all these nine data points are used when analyzing the inner kinematics
of the BCG, as described in the following sections.
\newline

Our X-ray analysis is based on archival Chandra ACIS-I observations (ObsId 4966, 18611, 18818). We have reprocessed them with a standard pipeline based on CIAO 4.9 \citep{CIAO} and CALDB 4.7.4 to create a new events-2 file which includes filtering for grade, status, bad pixels, and time intervals for anomalous background levels. We obtain a cumulative good time interval of 123.1 ksec. 
All the point sources detected with the CIAO routine {\it wavdetect} have been masked and not considered in the following analysis.
We estimate a local background for both the spatial and spectral analysis from two regions of 15 and 30 arcmin$^2$, respectively, located about 10 arcmin southward from the X-ray peak.
\newline

\citet{caminha16} determined the cluster total mass distribution near its center by using a model that include the position of
the cluster center as a free parameter. They found that the cluster center is nearly coincident with the position of the BCG, (see x and y parameter values in their Table 5) therefore we set the cluster center on the BCG at RA=$22^h48^m43.99^s$ and Dec=$-44^{\circ}31'50.98"$. 
We note that for all the profiles showed in Figure \ref{fig:mtot}, we assume the BCG as the cluster center. 

\section{Methodology} \label{s:mod}
We can characterize the total cluster mass profile $\mtot(r)$ as the sum of different components,  
\begin{equation}\label{eq:mpart}
 M(r) \equiv \mtot (r) = \mbcg(r) + \mgal(r)+ \mgas(r) + \mdm(r).
\end{equation}
The first three components account for the total baryonic mass, $\mbcg(r)$ is the stellar mass profile of the BCG, $\mgal(r)$ is the stellar mass profile of all the other cluster galaxies, and $\mgas(r)$ is the hot intra-cluster gas mass profile. The $\mdm(r)$ is the DM component of the total mass profile.  Hereafter, we describe how we determine the different mass components.

\subsection{The baryonic components}\label{ss:mbar}

Following \citet{sand04}, we model $\mbcg$ with a \citet{jaffe83} profile, 
\begin{equation}\label{eq:mbcg}
 \mbcg (r) \equiv M_{\mathrm{Jaf}} (r)= M_* \frac{r}{\rj} \frac{1}{(1+r / \rj)} ,
\end{equation}
where  $M_*$ is the total stellar mass of the BCG, and  $\rj$ is the characteristic scale radius of the model. The scale radius $r_j$ is determined by fitting a de Vaucouleurs model to the surface-brightness profile of the BCG, knowing that it is related to the effective radius by $ \rj = R_e / 0.76$ \citep{jaffe83}. The effective radius $R_e$ is defined as the radius that contains half the total BCG luminosity (computed as the Kron magnitude in the  R band, in our case). The BCG total stellar mass in Equation~\ref{eq:mbcg}, $M_*$, is obtained by fitting the spectral energy distribution (SED) and the MUSE spectrum assuming a Salpeter stellar initial mass function (IMF; see M20 for details). We find the effective radius $R_e= 17.0 \pm 1.6$ kpc \citep[see also table A1 in ][]{tor18}, that corresponds to $r_j = 22.7 \pm 2.1$ kpc, and the total stellar mass $M_* = 1.2 ^{+0.2}_{-0.6}  \times 10^{12} M_\odot$.  

The second baryonic component of $\mtot$ in Equation \ref{eq:mpart},  $\mgal$, is the stellar-mass profile of the other cluster member galaxies, BCG excluded. The stellar mass of each member galaxy is obtained by SED fitting, using the 5-band ground-based  photometry, based on observations with the WFI at the ESO 2.2m telescope \citep{gruen13}, after a cross-calibration with the HST photometric bands(see M20). On average, galaxies have their stellar masses estimated with an error $\Delta \log(M_*) =0.155 $ (see M20).
When evaluating $\mgal$, we consider all the galaxies with $M_* \geq 10^{10}M_\odot$ within the projected radius $0.03 \leq R \leq 2.75$ Mpc . The minimum radius has been chosen to exclude the contribution from the BCG, which is already accounted for in $\mbcg(r)$.  The maximum  radius is the radius within which our galaxy member sample is complete at the $80 \%$ level. The completeness in stellar mass has been computed from the completeness in magnitude following M20. We evaluate the projected stellar mass density profile by summing the total stellar mass of all galaxies in radial bins, weighted by the inverse of their completeness. The uncertainties on this profile have been derived with a bootstrap procedure on the galaxies in each radial bin,
adding in quadrature the bootstrap-derived variance and the mean error on the stellar masses ($\Delta \log(M_*) =0.155 $). We then deproject the 2D stellar mass profile with its uncertainties,
using the Abel inversion equation \citep{binney87}, and integrate the 3D density profile to obtain $\mgal(r)$. The gas mass profile in Equation \ref{eq:mpart}, $\mgas(r)$, is evaluated from the numerical integration over the cluster's volume of the gas density profile.

Assuming a spherical geometry, the gas density profile has been recovered from the deprojection of the X-ray surface brightness profile \citep[see e.g., Sect.~1.2 in][]{ettori13xmass}. This profile has been extracted from the exposure corrected image in the 0.7--2 keV energy band (where the S/N is maximized), which represents the effective area at each sky position of the detector and accounts for the dithering effects of the telescope.

\subsection{The Dark Matter component}
\label{ss:mdm}

Once the three baryonic components $\mbcg, \mgal, \mgas$ have been determined directly from the observations, as described in  Section \ref{ss:mbar}, in order to determine $\mdm(r)$, we seek a solution for the spherical Jeans equation of dynamical equilibrium,  
\begin{equation}\label{eq:jeans}
  \frac{\der \phi}{\der r} = \frac{G \mtot(r)}{r^2} = - \frac{\sigtr}{r} \left( \frac{\der \ln \nr}{\der \ln r} + \frac{\der \ln \sigtr}{\der \ln r} + 2\br\right),
\end{equation}
where $\mtot(r)$ is the total mass profile of Equation~\ref{eq:mpart}, $r$ is the three-dimensional cluster-centric distance, $\phi$ is the cluster potential, $\nr$ is the three-dimensional number or luminosity density profile of the tracer of the potential, $\sr$ its radial velocity dispersion profile, and $\br$ its anisotropy profile, $\br \equiv 1- \sigma_\theta^2(r)/\sigma_r^2(r)$, with $\sigma_\theta$ and $\sigma_r$ , respectively, the tangential and radial component of the velocity dispersion.  We solve the Jeans equation by fitting at the same time the BCG line-of-sight (LOS) velocity dispersion profile (see Figure \ref{fig:bcg}), and the projected phase-space distribution of the other cluster galaxies (see MAMPOSSt method below). While $\mtot(r)$ is common to both these tracers, $\nr$ and $\br$ are different for the two tracers of the potential. 

The baryonic components to $\mtot(r)$ are described in Section~\ref{ss:mbar}. As for the DM component, $\mdm$, we model it with a generalized NFW (gNFW) profile,
\begin{eqnarray}\label{eq:mdm}
\mdm (r)  &\equiv M_{\mathrm{gNFW}} (r) = \mtwodm \, (r / \rtwodm )^{3-\gdm}   \nonumber \\
&\times  \frac{_2F_1(3-\gdm,3-\gdm,4-\gdm,-\rtwodm / \rsdm)}
       {_2F_1(3-\gdm,3-\gdm,4-\gdm,-r / \rsdm)} 
,\end{eqnarray}
where $_2F_1(a,b,c,x)$ is the hypergeometric function \citep{weisstein98,mamon19}. This model is characterized by three parameters: the inner slope, $\gdm$, the scale radius, $\rsdm$, and the mass at $\rtwodm$, $\mtwodm$. The relation between $\rmt$, $\rs$ and $\gamma$ for the gNFW model is
\begin{equation}\label{eq:gamrs}
 \rmt = (2-\gamma) \rs,
\end{equation}so that $\rmt \equiv \rs$ for the NFW profile, where $\gamma = 1$.

When we trace the potential using the BCG LOS velocity dispersion, we identify $\nr$ with the luminosity density profile, $\nu_{BCG}(r)$, that we model with a Jaffe profile,
\begin{equation}\label{eq:jaffedens}
\nu_{\mathrm{BCG}}(r) \propto (r/\rj)^{-2} \, (1+r/\rj)^{-2}.
\end{equation}
We note that we do not need to consider the normalization of $\nu_{BCG}(r)$ since it enters Equation~\ref{eq:jeans} only with its logarithmic derivative. The value of $\rj$ is obtained by  fitting the surface-brightness profile of the BCG as explained in Section~\ref{ss:mbar}.
For the anisotropy profile $\br$ of the BCG stars, we adopt the \citet{osipkov79} and \citet{merritt85} model, 
\begin{equation}\label{eq:betaom}
    \beta_{\mathrm{OM}}(r) = r^2 /(r^2+\ra^2),
\end{equation}
with the scale radius $\ra$ as a free parameter. 

For the other tracer of the potential that we consider, that is the cluster member galaxies, we identify $\nr$ with the number density profile that we model with a NFW profile,
\begin{equation}\label{eq:nugal}
    \nu_{\mathrm{gal}}(r) \propto (r/\rn)^{-1} \, (1+r/\rn)^{-2},
\end{equation}
where the only parameter is $\rn$, since the normalization of the profile is irrelevant in Equation~\ref{eq:jeans}. For the anisotropy profile $\br$ of the cluster member galaxies, we consider three different models,
\begin{enumerate}
\item constant velocity anisotropy with radius, $\br=\binf$, 
\item a simplified version of the model of \citet{tiret07} $\br \equiv \beta_{\mathrm{T}}(r)= \binf \, r/(r+\rmt)$ where the orbits of the galaxies are isotropic in the inner part of the cluster and radial in the outer part, 
\item the \virgo{opposite} model from \citet{biviano13}, $\br \equiv \beta_{\mathrm{O}}(r)=\binf \, (r-\rmt)/(r+\rmt)$, that allows orbits to be tangential in the inner part and radial in the outskirts, or viceversa. 
\end{enumerate}
We note that there is only one free parameter in all three anisotropy models, $\binf$, forced to be $<1$ to avoid unphysical solutions, while $\rmt$ is the same scale radius of Equation~\ref{eq:gamrs}. 

In summary, to determine $\mdm(r)$ we must constrain the following  parameters,
$\gdm, \rsdm, \rtwodm, \rn, \binf, \ra$, and we do this by combining the likelihoods that we obtain from the BCG internal stellar dynamics, $\likbcg$,  and the projected phase-space distribution of the cluster members, $\likgal$.

To compute $\likbcg$, we model the observed projected LOS stellar velocity dispersion profile of the BCG as:
\begin{equation}\label{eq:sig2D}
\sigma_{\mathrm{p,BCG}}^{2}(R) = \frac{2}{(M_{*}/L)\;\Sigma_*(R)} 
\int_R^{\infty} \der r' \left[1-\frac{R^{2}}{\ra^{2}+r'^{2}}\right]\frac{\nu_{\mathrm{BCG}}(r') \sigma_{r}^{2}(r') r'}{(r'^{2}-R^{2})^{1/2}},
\end{equation}
where $\sigma_{r}$ is the radial velocity dispersion profile, and where we have used the chosen velocity anisotropy profile of equation~\ref{eq:betaom}. The radial velocity dispersion profile $\sigma_{r}(r)$ is related to the total mass profile $\mtot(r)$ via:
\begin{equation} \label{eq:sig3D}
\sigma_{r}^{2}(r) = \frac{G \int_r^{\infty} \der r' \nu_{\mathrm{BCG}}(r') \mtot(r') \frac{\ra^{2}+r'^{2}}{r'^{2}}}{(\ra^{2}+r^{2}) \nu_{\mathrm{BCG}}(r)}.
\end{equation}
The likelihood $\likbcg$ follows from a $\chi^2$ ($\likbcg \propto \exp^{-\chi^2/2}$) comparison between the model $\sigma_{p,BCG}$ of Equation~\ref{eq:sig2D} and the observed line-of-sight BCG stellar velocity dispersion (see Figure~\ref{fig:sigmaprof}). This likelihood depends on the free parameters of $\mdm$, (and thus $\mtot(r)$) $\gdm, \rsdm, \rtwodm$ in our case, and on the free parameter of  $\beta_{\mathrm{OM}} (r)$, $\ra$. 

\begin{figure}
 \includegraphics[width=0.50\textwidth]{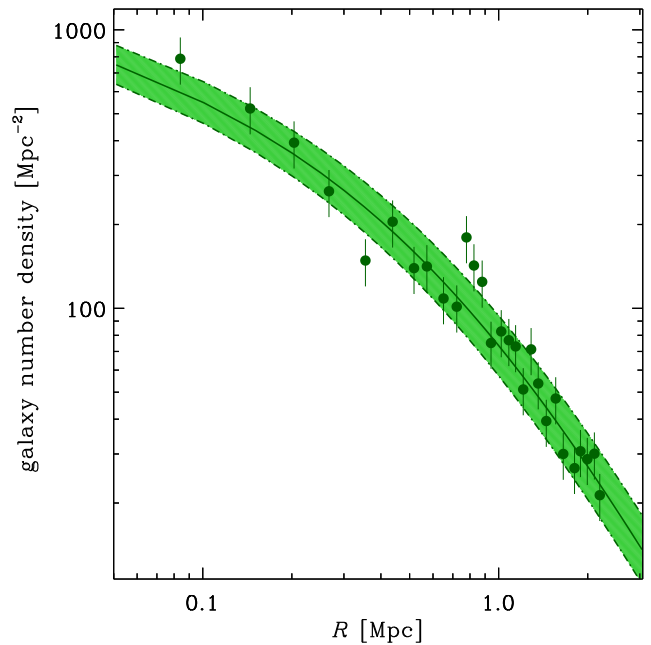}
\caption{Projected galaxy number density profiles $n(R)$ (points with 68 \% error bars), for the member galaxies, corrected for the incompleteness. The solid line represents the best-fit projected NFW model with 68 \% errors (shaded area). }
\label{fig:den}
\end{figure}

To obtain the likelihood $\likgal$, we use the \texttt{MAMPOSSt} technique of \citet{mamon13}. Given the models for $\mtot(r)$ and $\br$, \texttt{MAMPOSSt} estimates the best-fit parameters ($\gdm, \rsdm, \rtwodm, \binf$ in our case) that maximize the probability of finding the cluster member galaxies at their observed positions in projected phase-space. 
In the \texttt{MAMPOSSt} analysis we only consider the subsample of 792 cluster members in the radial range 0.05--2.36 Mpc, to exclude the BCG at the very centre (since it is considered separately in our analysis), and to limit the analysis to the virial region of the cluster where the Jeans equation is valid. As a first estimation of he virial region we adopt $2.36$ Mpc, according to the lensing analysis (see Section \ref{s:res}),however we checked that this choice does not affect our results. We run \texttt{MAMPOSSt} in the so-called split mode, by separately fitting  the spatial and velocity distribution of cluster members. \citet{mamon13} have shown that this is equivalent to a simultaneous fit to the two distributions. The split mode allows us to deal with the radial incompleteness of the spectroscopic sample in a simple way, as detailed hereafter. The observed profile is corrected for incompleteness (see Section \ref{s:data}). To take into account the errors on the the completeness values, we perform  a Monte-Carlo analysis. In each radial bin we randomly draw 10000 values from a Gaussian distribution centered on the mean completeness with a sigma corresponding to the completeness uncertainty (see M20). We calculate the characteristic radius of $\nu_{gal}$ (Equation~\ref{eq:nugal}), $\rn$,  by a maximum likelihood fit to the observed projected number density profile of cluster galaxies, using the projected NFW model \citep{Bartelmann96}. We also considered the \citet{king62} (cored) model, but we found that the projected NFW model provides a higher likelihood. For the NFW model, we find a best fit value of $ \rn=0.76^{+0.08}_{-0.07} \; \mathrm{Mpc}$. In Figure \ref{fig:den}, we show the observed galaxy projected number density profile, $\nu_{gal}$, and its best-fit projected NFW model.

Having determined $\rn$, the remaining free parameters in the combined dynamical analysis of the BCG velocity dispersion profile and of the cluster galaxy velocity distributions are: 
\begin{equation}\label{eq:dmpar}
\mathbf{\theta_M} \equiv \{\gdm, \rsdm, \rtwodm, \binf, \ra \}.
\end{equation}
We implemented our full procedure in the COSMOMC code \citep{cosmomc}. We determined the best-fit values of $\mathbf{\theta_M}$ by searching for the maximum of the sum $\ln \likgal+\ln \likbcg$, and determined the confidence intervals of $\mathbf{\theta_M}$ via a MCMC (MonteCarlo Markov Chain) procedure. 

\section{Results} \label{s:res}

\begin{figure*}
 \includegraphics[width=0.90\textwidth]{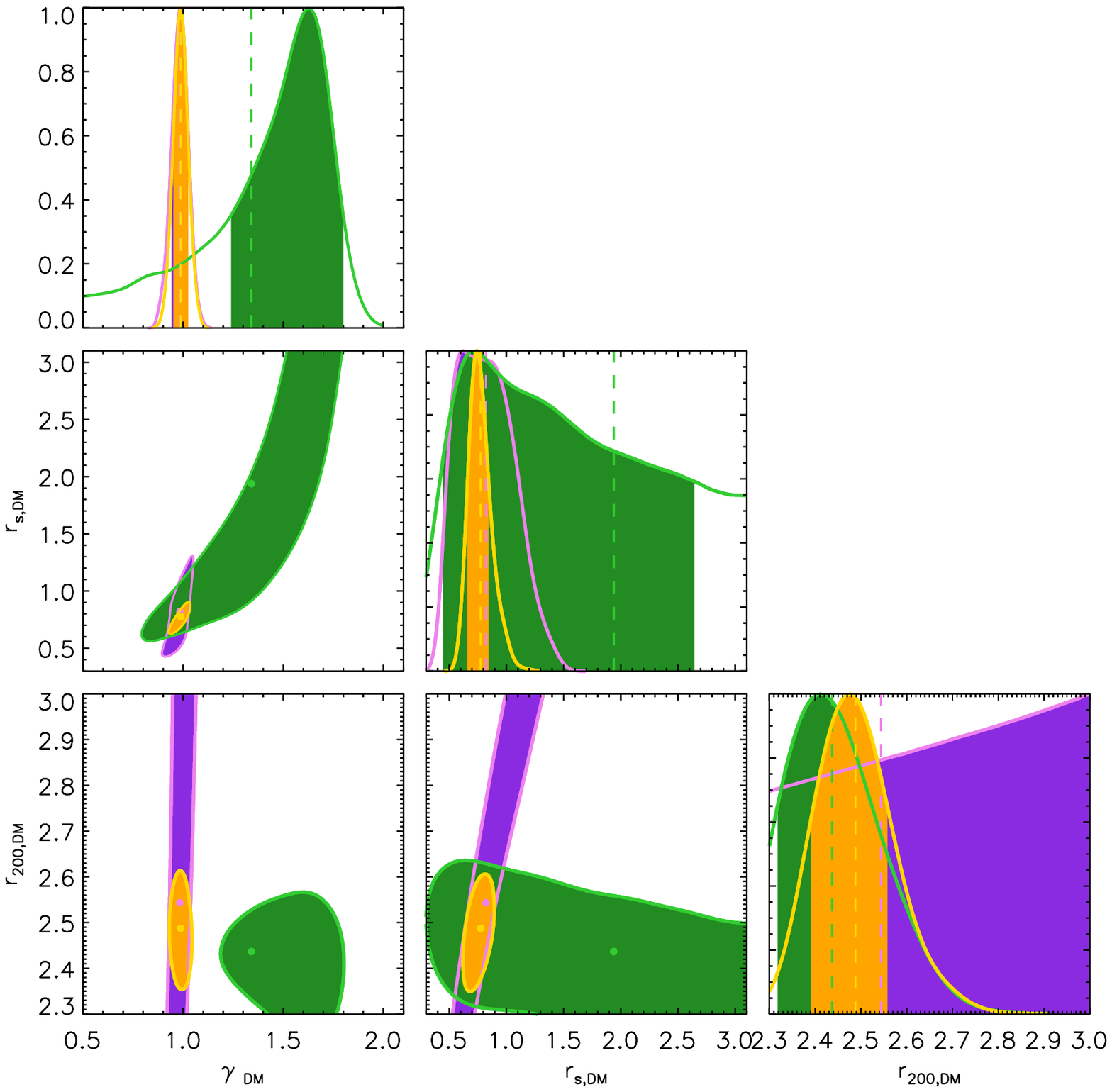}
\caption{Constraints on $\mtot(r)$ parameters $\gamma,\rs,\rtwo$ at the 68 \% confidence level, as obtained after marginalization over the remaining parameters. Violet, green, and orange colors indicate the constraints obtained from $\likbcg, \likgal$, and $\likbcg+\likgal$, respectively.}
\label{fig:galstar_tri}
\end{figure*}

\begin{table*}
\centering
\caption{Best-fit free parameters include in the $\mtot(r)$ model (see Equation \ref{eq:mpart}) and their $68 \%$ errors.}
\begin{tabular}{|l|c c c c c c|}
\hline
Tracer & $\mdm$ & $\rtwodm$ & $\rsdm$ & $ \gdm$ & $\binf$ & $\ra$ \\
 &$10^{15} [M_\odot]$ & [Mpc] & [Mpc] &  &  &  [Mpc]\\
 \hline
       Galaxies   &$ 2.37   \pm0.32 $ & $  2.44   _{- 0.12  }^{+ 0.09}$ & $ 1.94   _{- 1.52  }^{+ 0.72}$ & $ 1.34   _{- 0.12  }^{+ 0.46}$ & $ 0.80   _{- 0.21  }^{+ 0.10 }$ &       -   \\
            BCG   &$ 2.69   \pm0.98 $ & $  2.54   _{- 0.16  }^{+ 0.46}$ & $ 0.82   _{- 0.30  }^{+ 0.19}$ & $ 0.98   _{- 0.04  }^{+ 0.04 }$ &   - & > 1   \\
   Galaxies+BCG   &$ 2.52   \pm0.26 $ & $  2.49   _{- 0.10  }^{+ 0.08}$ & $ 0.78   _{- 0.12  }^{+ 0.08}$ & $ 0.99   _{- 0.04  }^{+ 0.04}$ & $ 0.75   _{- 0.19  }^{+ 0.07}$ & >1   \\

\hline
\end{tabular}
\label{tab:par}
\end{table*}

In Figure \ref{fig:galstar_tri}, we show the constraints on the mass profile parameters that we obtain by combining the likelihood from the analysis of the cluster galaxy dynamics, $\likgal$, and the likelihood from the stellar dynamics of the BCG, $\likbcg$. We show the 2D constraints at the 68\% confidence level for the combinations of ${\gdm,\rsdm,\rtwodm}$ parameters that characterize $\mdm(r)$ (see Equation~\ref{eq:mdm}), as obtained after marginalizing over the remaining parameters (included $\binf$ and $\ra$). We use different colors to distinguish the constraints that we obtain from $\likgal$ and $\likbcg$ individually (in green and purple, respectively) and from the two likelihoods combined (in orange). Using $\likgal$ alone, the value of $\rtwodm$ is well-constrained while there is a strong degeneracy in the values of the two parameters $\gdm$ and $\rsdm$. Clearly, this degeneracy indicates that we cannot constrain the inner behavior of $\mdm(r)$ from galaxy kinematics alone.  On the other hand, as expected, using the information from  $\likbcg$ alone, it is not possible to constrain $\rtwodm$, while the inner slope $\gdm$ is well-constrained. By combining $\likgal$ and $\likbcg$, we obtain the values of the  $\mathbf{\theta_M}$ parameters (Equation \ref{eq:dmpar}) and their 68 \% marginalized errors listed in Table \ref{tab:par}. For each parameter in Table \ref{tab:par}, we list the mean and the 68 \% error of the distribution as obtained after marginalized over all the others parameters. These values are obtained for the \citet{tiret07} $\beta_T$ profile, that provides the highest likelihood among the three considered anisotropy models for the cluster galaxies (see Section~\ref{ss:mdm}). The main result of this paper is that the value of the inner slope of the DM profile, as obtained from our full dynamical analysis, is $\gdm = 0.99   \pm  0.04 $,  which is in perfect agreement with what expected from simulations (see Section \ref{s:disc} for the discussion).

\begin{figure}
\begin{center}
\begin{minipage}{0.5\textwidth}
  \resizebox{\hsize}{!}{\includegraphics{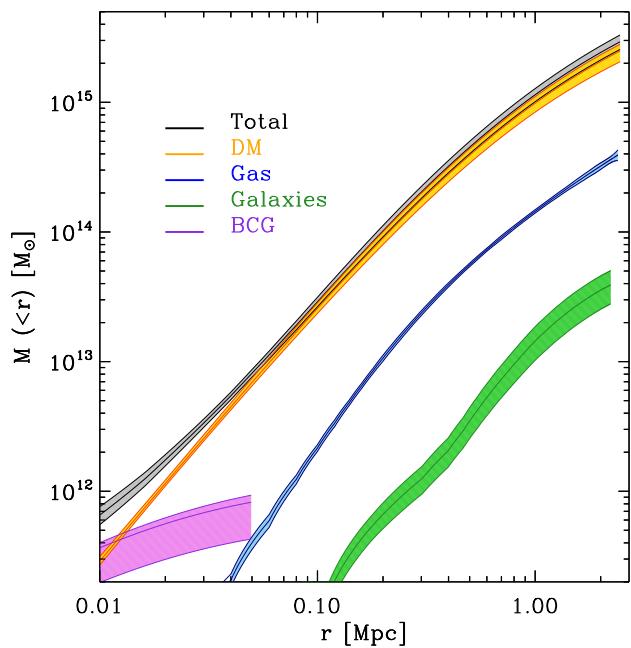}}
\end{minipage}
\end{center}
\caption{3D mass profiles of the different cluster components, as derived from our analysis. Black: the total mass $\mtot(r)$; orange: the DM component $\mdm(r)$; violet: the stellar mass of the BCG $\mbcg(r)$; blue: the hot intra-cluster gas $\mgas(r)$; and green: the stellar mass of the other cluster galaxies $\mgal(r)$. The  shaded areas show the $68 \%$ confidence regions.}
\label{fig:mpart}
\end{figure}

\begin{figure}
\begin{center}
\begin{minipage}{0.5\textwidth}
  \resizebox{\hsize}{!}{\includegraphics{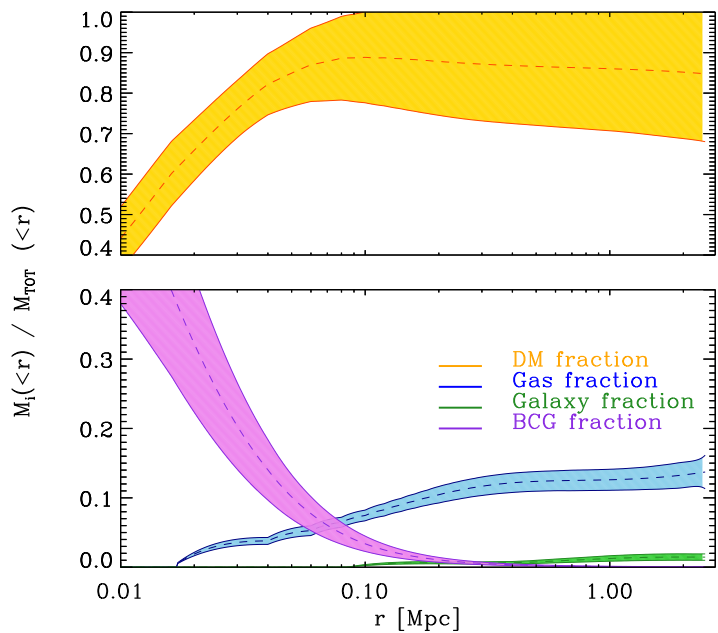}}
\end{minipage}
\end{center}
\caption{
Ratios  between  the  3D mass  profiles  of  the  different  cluster  components  and  the  total 3D mass  profile  within  a  given  radius.
In orange, $\mdm(r)$ / $\mtot(r)$; in violet, $\mbcg(r)$ / $\mtot(r)$; in blue, $\mgas(r)$ / $\mtot(r)$; and in green, $\mgal(r)$/ $\mtot(r)$. The  shaded areas show the $68 \%$ confidence regions.}
\label{fig:mpartrat}
\end{figure}

The various mass profiles contributing to $\mtot(r)$ (see Equation~\ref{eq:mpart}) are shown in Figure~\ref{fig:mpart}. In black, we show $\mtot(r)$ as obtained by simply adding the different mass contribution at each radii. The contribution of each component to the total mass is shown in Figure~\ref{fig:mpartrat}. 
The  stellar mass profile of the BCG, $\mbcg(r)$, is shown with violet curves. Its contribution to $\mtot(r)$ is the dominant one only at the very centre, and at $r \gtrsim 15$ kpc. As for the other baryonic components, the stellar mass profile of all the other cluster galaxies, $\mgal(r)$, is shown in green, while the hot intra-cluster gas mass profile,$\mgas(r)$, is shown with blue curves. 
Of these baryonic components, $\mgal(r)$ makes a negligible contribution to $\mtot(r)$ at any radius, while $\mgas(r)$, although negligible near the center, contributes significantly to $\mtot(r)$ at increasingly larger radii.  The stellar component (BCG excluded) provides only $1.5 \pm 0.4 \% $ of the total amount of mass to the cluster, and the hot gas the $\sim 13 \pm 3 \%$ at $r_{200}$ and thus the DM contributes at $\sim 84 \pm 14 \%$ level at the same radius. These values are consistent with what has been found by \citet{annunziatella17} for the MACS J0416.1-2403 cluster of comparable total mass and redshift. \citet{bivsal06} found a fraction of hot gas mass that increase from 7 to 11 \% from $0.1 r_{100}$ to $r_{200}$, for the galaxy stellar mass from 4\%\ to 2\% in the same radial range. 
Such results come from the analysis of the stack of 59 nearby clusters with a final mass of   $\sim 6 \times 10^{14} M_\odot$ from the ESO Nearby Abell Cluster Survey \citep{Katgert+96}.
\citet{eckert19} computed the hot gas mass fraction out to $\rtwo$ for the {\it XMM-Newton} Cluster  Outskirts  Project \citep[X-COP,][]{eckert17} sample of 12 nearby clusters with a mass range $M_{200} \sim [0.6-2.] \times 10^{15} M_\odot$. 
At $\rtwo$, they measured a median value of $f_{gas}=0.149^{+0.009}_{-0.008}$.
This value exceeds the reference ``universal'' gas fraction of $f_{gas, u} = 0.134 \pm0.007$ evaluated as $f_{gas,u} = Y_b \Omega_b/\Omega_m -f_*$,
where $\Omega_b/\Omega_m = 0.156\pm0.003$ is constrained by the CMB power spectrum in \citet{planck15}, $Y_b$ is the baryon depletion factor predicted from hydrodynamical simulations and $f_*=0.015\pm0.005$ is obtained from a compilation of recent results on the stellar mass fraction \cite[see][for details]{eckert19}.
Similarly, we can estimate the baryon depletion factor, $Y_b$, representing the fraction of the baryons enclosed within a given radius, for \cl\ and compare it with published results.
In hydrodynamical simulations, $Y_b$ in cluster's halos is expected to be 85-95 \% \citep{planelles13,eckert19}. 
In \cl, we measure a value of $Y_b = 0.92 \pm 0.21$ at $\rtwo$ in agreement with previous constraints, mostly from numerical simulations.
 
\begin{figure}
\begin{center}
\begin{minipage}{0.5\textwidth}
 \resizebox{\hsize}{!}{\includegraphics{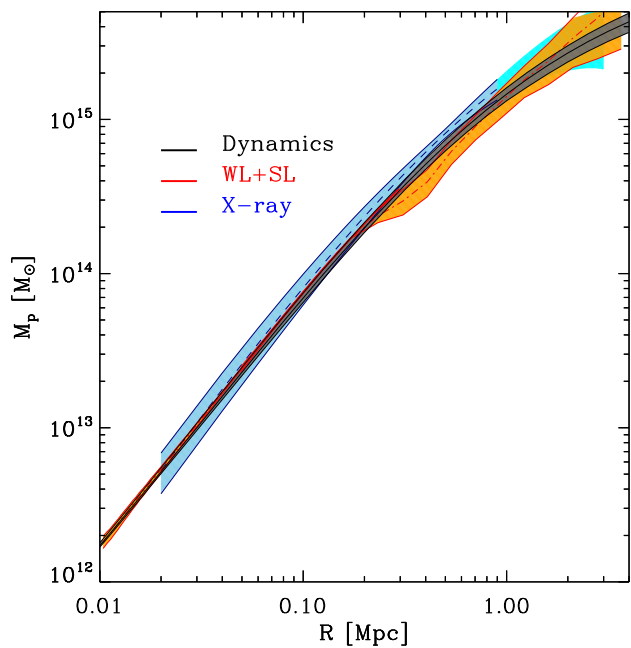}}
\end{minipage}
\end{center}
\caption{Total projected mass profile comparison. Black: projection of the 3D total mass in Figure \ref{fig:mpart} as derived from our dynamical analysis. Red: result from the combined weak lensing 
analysis by \citep{umetsu16} and strong lensing analysis by \citet{caminha16}. Blue: result from the X-ray data analysis using the hydrostatic equilibrium equation; in cyan we indicate the region where the X-ray profile has been extrapolated. The shaded areas show $68 \%$ confidence regions.}
\label{fig:mtot}
\end{figure}

In Figure \ref{fig:mtot}, we compare the total mass profile as derived from different analyses. For the sake of a better comparison with the profiles derived from the gravitational lensing analyses, we compare here projected mass profiles. Our derivation of $\mtot(r)$ from the dynamical analysis, after
Abel projection, is shown in black.
The blue curves represent the mass profile obtained from the Chandra X-ray data. This profile has been recovered by solving the equation of hydrostatic equilibrium assuming spherical symmetry,
\begin{equation}\label{eq:hydro}
M_{\mathrm{x-ray}}^{\mathrm{tot}} (r) = -\frac{kT_{\mathrm{gas}}(r) r}{G \mu m_p} \left( \frac{\partial \ln n_{\mathrm{gas}} }{\partial \ln r } + \frac{\partial \ln T_{\mathrm{gas}} }{\partial \ln r } \right) 
,\end{equation}
where $G$ is the gravitational constant, 
$\mu=0.59$ is the mean molecular weight in a.m.u., 
$m_p$ is the proton mass, $k$ is the Boltzmann constant $T_{gas}(r)$ and $n_{gas}(r)$ are the gas temperature and density profiles, in 3D. The gas density profile has been obtained from the geometrical deprojection of the X-ray surface brightness profile resolved up to 2 Mpc. The gas temperature profile is obtained by the spectral analysis performed on 16 independent spatial bins up to $\sim 900$ kpc, by requiring $\sim$3000 net counts in the 0.5--7 keV band in each bin. The spectral fitting has been done in Xspec 12.9.1 \citep{arnaud96}, using a thermal component ({\tt apec}) with three degrees of freedom (normalization, temperature, and metallicity), absorbed by a Galactic column density $n_H$ fixed to the local value of $1.24 \times 10^{20}$ cm$^{-2}$ \citep{lab05}, and at the nominal redshift of 0.3458. A NFW functional form for the gravitational potential is adopted with two free parameters (normalization and $R_{200}$). These parameters are constrained by performing a grid-based search for a minimum in the distribution of the $\chi^2$ evaluated by comparing the observed spectral temperature profile (with the propagated relative errors) and that predicted from the inversion of the hydrostatic equilibrium equation in which the observed gas density profile and the assumed mass model are used. The predicted temperature profile is then projected in each annulus of the spectral analysis \citep[for further details see][]{ettori10}. At each radius, we associate a symmetric error on the mass profile that represents the range of values allowed from the $1 \sigma$ statistical uncertainties on the 2 free parameters (i.e., as defined from the region enclosed within $\Delta \chi^2=2.3$). 

The red curves in Figure (\ref{fig:mtot}) represent the total mass profile as obtained from the combination of the weak and strong lensing analyses. Here we have improved upon the earlier CLASH lensing work of \citet{umetsu16} , by combining the wide-field weak lensing data of \citet{umetsu14} with the latest strong lensing mass model of \citet{caminha16}.  The weak lensing mass profile of \cl\ is based on ESO/WFI data and presented in \citet{umetsu14}, who presented a wide-field, weak-lensing shear and magnification analysis of 20 CLASH clusters. \citet{umetsu16} performed a joint weak+strong lensing analysis of the CLASH sample, in combination with 16-band HST observations \citep{zitrin15}. An improved strong lensing analysis of AS1063 has been carried out by \citet{caminha16} using 16 background sources (with $1.0 \le z \le 6.1$) that are multiply lensed into 47 images, 24 of which are spectroscopically confirmed and belong to ten individual sources. The data-set used for their strong lensing analysis comprises the CLASH imaging data and the spectroscopic follow-up observations, with the VIMOS and MUSE on the VLT.

The three determinations of the total cluster mass profile are in good agreement. The agreement in the inner slope of the profile between our dynamical analysis and the strong lensing determination is particularly remarkable. The only relevant difference among the three  is present in the weak lensing profile at $R\sim 0.3-0.6$ Mpc. Perhaps this discontinuity is related to a substructure at $\sim 0.5$ Mpc from the cluster centre, a residual of a recent off-axis merger, as indicated by the analysis of \citet{gomez12}. This merger event is also suggested by the elongated shape of the X-ray emission \citep{caminha16}, along the same direction of a large-scale galaxy filament \citep{melchior15}. We note that the total mass of the cluster out to $\rtwotot = 2.63 \pm 0.09$[Mpc] as obtained  from the dynamical analysis of the cluster galaxies is $\mtot = 2.87 \pm 0.3 \times 10^{15} [M_\odot]$  by assuming a NFW model \citep[see][]{biviano13}, which is consistent with the DM mass fraction shown in Figure \ref{fig:mpartrat} and Table \ref{tab:par}. By fitting the NFW model to the lensing mass profile we find $\rtwotot = 2.36 \pm 0.23$[Mpc] and thus $\mtot = 2.17 \pm 0.6 \times 10^{15} [M_\odot]$, in agreement with results from the dynamical analysis. By extrapolating the best-fit NFW mass model, the hydrostatic mass from X-ray analysis at these $r_{200}$ is $3.7 \pm 0.6 \times 10^{15} [M_\odot]$ and $4.2 \pm 0.8 \times 10^{15} [M_\odot]$ at 2.36 and 2.63 Mpc, respectively.

\begin{figure}
 \includegraphics[width=0.5\textwidth]{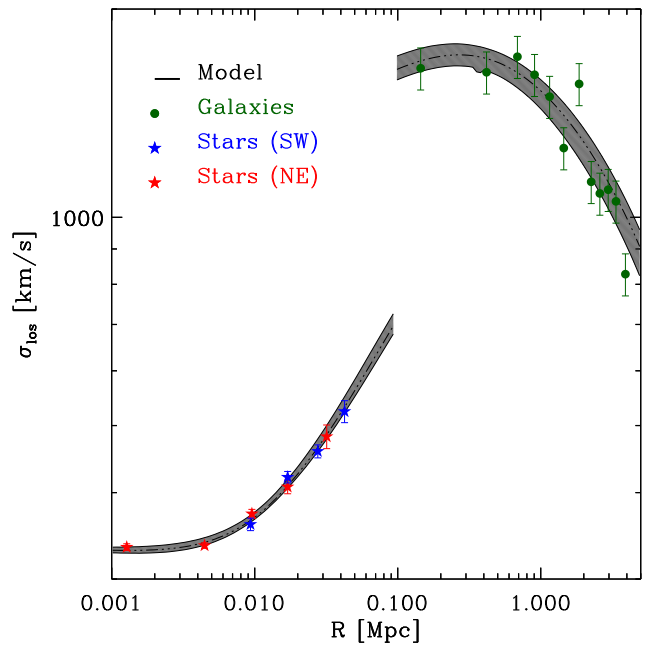}
\caption{LOS velocity dispersion of the two probes of the cluster potential well, BCG stars (red and blue stars as in Figure \ref{fig:sigmaprof}) and cluster galaxies (green points), as a function of the projected cluster-centric distance.  Error bars refer to a 68 \% confidence level. The black curve represents the LOS velocity dispersion obtained for the best-fit mass model derived from the MCMC analysis, which includes the combined likelihood $\likbcg+\likgal$, from the dynamical analysis of the BCG stellar component and member galaxies (Table \ref{tab:par}). The  shaded area shows the $68 \%$ confidence region.
} \label{fig:bcg}
\end{figure}

We show in Figure~(\ref{fig:bcg}), the velocity dispersion profile of the two probes we use in our analysis: the LOS velocity dispersion of the cluster member galaxies and the LOS velocity dispersion of the stellar component of the BCG. Derivation of the latter has been described in Section~(\ref{s:data}). As for the LOS velocity dispersion of the cluster member galaxies, we use the bi-weight estimator \citep{BFG90} applied to the velocity distribution of the galaxies in each radial bin, and we define the radial bins in such a way as to keep the same number of galaxies in each bin. In the same figure, we show the best-fit model velocity dispersion and its 68 \% confidence interval, as obtained from the combined MCMC likelihood analysis, $\likbcg+\likgal$ (see Table \ref{tab:par}). To calculate the LOS velocity dispersion profiles, we use Equations~\ref{eq:sig3D} and \ref{eq:sig2D} for the LOS velocity dispersion profile of the BCG and the equivalent ones for the LOS velocity dispersion profiles of cluster galaxies \citep[Eqs. 9 and 30 in][]{mamon13}. We stress that the gap in the LOS velocity dispersion profile is just due to the fact that we are showing the velocity dispersion profiles of two different tracers, here, BCG stars at $R\lesssim 0.1$ Mpc and cluster galaxies at $R\gtrsim 0.1$ Mpc.  The two tracers have different density profiles so they are not expected to have the same velocity dispersion profile. In particular, the density profile of the stars drop as $r^{-4}$ at large distances from the BCG center, while the density profile of galaxies drop as $r^{-1}$ at small distances from the cluster center. Since the logarithmic derivatives of the density profiles of the two tracers are very different, the velocity dispersions also differ.

\section{Discussion}\label{s:disc}

By using the kinematics of the \cl\ cluster galaxy members and the stellar velocity dispersion profile of the BCG, we have constrained the inner slope parameter of the cluster DM profile (modeled as a gNFW), $\gamma=0.99 \pm 0.04$ (1 $\sigma$ uncertainty). This value is compatible with, and in fact almost identical to, the expected value for the NFW model. While the NFW model was first derived for halos identified in DM-only cosmological simulations \citep{NFW96}, subsequent analyses have shown that the NFW model is also a good description of the DM profile of halos identified in cosmological simulations that include baryons \citep{Schaller+15,He19}. It has been argued that, if halos grow by mergers that produce significant mixing between stars and DM, it is the total mass, not the DM density profile that displays the NFW slope near the center \citep{LW15}. However, even in the absence of significant mixing between stars and DM, simulations indicate that massive halos are dominated by DM down to $<0.01 \, r_{200}$ \citep{Schaller+15}, which is in agreement with our findings (see Figure~\ref{fig:mpart}). To probe these very inner regions of massive clusters, a study of the BCG kinematics is essential since it is uncommon to find strong lensing features at distances $<10 $ kpc from the cluster center.

Several DM alternatives to CDM predict cored cosmological halos: self-interacting DM \citep{YSWT00,SS00}, axion DM \citep{MP15}, warm DM \citep{CARV00},  eventually combined with a  CDM component coupled to Dark Energy \citep{MMPB15}. Our result, taken at face value, is therefore supportive of a standard, collisionless CDM scenario. However, to fully understand whether $\gamma=1$ rules out alternative DM models, simulations with non-standard DM and baryons are required. Even if the DM is cold and collisionless, the inner slope of the halo DM density profile can be modified by baryonic processes. The inner slope can be flattened by AGN feedback heating and expelling gas from the central cluster region \citep{RFGA12,MTM13,Peirani+17}, by dissipationless stellar accretion onto the BCG \citep{Laporte+12}, and dynamical friction transferring energy from baryons to DM \citep{ElZant+04}. On the other hand, adiabatic contraction has the opposite effect of steepening the inner slope of the DM profile \citep{Blumenthal+86,Gnedin+04}. 

Other observations have found that the value of the very inner slope of the DM density profile of clusters is significantly smaller than 1 \citep{newman13b}, even if the slope of the total mass density profile is $\approx 1$ \citep{newman13a}. Our determination of $\gdm$ is quite different from the values found by \citet{newman13b} for other clusters of similar mass and at similar redshifts. The mean $\gdm$ value for the seven clusters analyzed by \citet{newman13b} is 0.54, with a dispersion of 0.20. If these values are typical of the population of clusters, \cl\ is an outlier at just slightly more than 2~$\sigma$.

\citet{Schaller+15} have argued that the $\gdm$ values found by \citet{newman13b} are biased to the lower end either because of their assumption of a Salpeter initial mass function (IMF) for the BCG stellar mass or because of their assumption of isotropic stellar orbits. However, we found $\gdm \approx 1$ even if we adopted a Salpeter IMF in our analysis, \citep[as seems to be appropriate for massive early-type galaxies, e.g., ][]{Treu+10,grillo10,grillo08}, and our best-fit solution for the velocity anisotropy of the BCG indicates (nearly) isotropic orbits (see Table~\ref{tab:par}). Moreover, \citet{NET15} find that groups have NFW-like DM density profiles with little dependence on the assumed orbital anisotropy. In their paper \citet{He19} fit
mock stellar kinematics and lensing data generated from the Cluster-EAGLE simulations, following the same approach of \citet{newman13b}, and retrieved
the inner density slopes of both the total and the dark matter mass distributions. \citet{He19} have suggested another possible origin of the low $\gdm$ values of \citet{newman13b}, namely an under-estimate of $\rsdm$ coupled to the error covariance between $\rsdm$ and $\gdm$.

Unless the $\gdm$ values found by \citet{newman13b} are indeed biased low, as suggested by \citet{Schaller+15} and \citet{He19}, our result indicates that there is significant cosmic variance in the values of $\gdm$ for clusters. Significant variance can result when clusters are observed in different dynamical states and at different ages because of the dynamical interplay between baryons and DM. An early assembly history corresponds to higher concentration of the halo density profile \citep{WT12}. A higher concentration may also appears as a consequence of a recent major merger, but only for a very short time, as the accreted subhalo passes to its orbital pericenter \citep{Klypin+16}. On longer timescales, unrelaxed halos have less concentrated mass density profiles than relaxed ones \citep{Neto+07}. The observed inner slope of the DM profile can also depend on the phase and strength of the central AGN feedback that change with time and with the mass ratio of the BCG to its parent halo \citep{Peirani+17}. 
 
 We need to extend our analysis to other clusters with similar high quality data to constrain the mean and the variance of $\gdm$. This will either disprove the observational results of  \citet{newman13b} or prove \cl\ to be a rather exceptional cluster.
 
\section{Conclusions}
\label{s:conc}
We carried out an accurate, full dynamical reconstruction of the mass density profile of the massive, $z=0.3458$ cluster \cl\, over the radial range $\sim 1$ kpc to $r_{200}$ ($\sim 2.7$ Mpc). Our determination is based on the solution of the Jeans equation for dynamical equilibrium as obtained from the application of an upgraded MAMPOSSt maximum likelihood technique \citep{mamon13} to the velocity dispersion profile of the BCG and to the velocity distribution of cluster members. The likelihood sampling was obtained with the MCMC method. The BCG velocity dispersion profile was measured with high precision, using MUSE integral field spectroscopy observations, while a very large sample of velocities of cluster members was obtained from a combination of the MUSE and VIMOS spectroscopic campaigns at the VLT (Mercurio et al. in prep). To disentangle the DM profile from the total mass profile we also made use of independent determinations of the stellar mass profile of cluster members (other than the BCG) and of the intra-cluster gas mass profile (based on Chandra X-ray observations). We note that these two latter components are negligible near the cluster center, dominated instead by the BCG and DM components.

We accurately determined the inner logarithmic slope of the DM density profile, modeled as a gNFW, $\gdm=0.99 \pm 0.04$, after marginalization over other free parameters of the mass and velocity anisotropy profiles. Our result is in full agreement with predictions from CDM/$\Lambda$CDM cosmological simulations \citep{NFW96,He19}, but inconsistent at $\sim 2 \sigma$ level with the distribution of $\gdm$ values found by \citet{newman13b} for clusters of similar mass and at similar redshifts. 
We derived the contribution of the different component of the mass profile, with their relative contribution to the total cluster mass (Figure \ref{fig:mpart}).
Furthermore, we compared the total mass profile as obtained from the dynamical analysis, with the one based on strong and weak lensing techniques, and the X-ray hydrostatic method: we find an excellent agreement among the projected masses at percent level within 1 $\sigma$ (Figure \ref{fig:mtot}), indicating a negligible hydrostatic bias.   

We plan to extend our analysis to other clusters of the CLASH-VLT data-set (Rosati et al. in prep) with additional, high-quality MUSE data in order to confirm the consistency of $\gamma$ with predictions from $\Lambda$CDM simulations, as well as to possibly determine the variance in the $\gdm$ values of clusters that may relate them to the properties of the clusters themselves and their BCGs.

\begin{acknowledgements}
This project is partially funded by PRIN-MIUR 2017 WSCC32 and INAF MainStream 1.05.01.86.20.
S.E. acknowledges financial contribution from the contracts ASI 2015-046-R.0 and ASI-INAF n.2017-14-H.0.
B. S. and M.G.  acknowledge the support from the grant MIUR PRIN 2015
\virgo{Cosmology and Fundamental Physics: illuminating the Dark
Universe with Euclid}.
M.G. acknowledges financial support from the University of Trieste
through the program \virgo{Finanziamento di Ateneo per progetti di ricerca
scientifica - FRA 2018}.
K.U. acknowledges support from the Ministry of Science and Technology
of Taiwan (grant MOST 106-2628-M-001-003-MY3) and from the Academia
Sinica Investigator Award (grant AS-IA-107-M01).
\end{acknowledgements}
\bibliography{libreria}

\end{document}